\documentstyle[12pt]{article}
\newcommand{\be}{\begin{equation}}
\newcommand{\ee}{\end{equation}}
\newcommand{\al}{\mbox{$\alpha$}}
\newcommand{\als}{\mbox{$\alpha_{s}$}}
\newcommand{\s}{\mbox{$\sigma$}}

\newcommand{\bi}[1]{\bibitem{#1}}
\newcommand{\fr}[2]{\frac{#1}{#2}}

\newcommand{\gm}{\mbox{$\gamma_{\mu}$}}
\newcommand{\Pm}{\mbox{$P_{\mu}$}}
\newcommand{\Pn}{\mbox{$P_{\nu}$}}

\newcommand{\ph}{\mbox{$\hat{p}$}}
\newcommand{\Ph}{\mbox{$\hat{P}$}}
\newcommand{\Gh}{\mbox{$\hat{\Gamma}$}}
\newcommand{\qh}{\mbox{$\hat{q}$}}
\newcommand{\kh}{\mbox{$\hat{k}$}}
\newcommand{\gn}{\mbox{$\gamma_{\nu}$}}

\newcommand{\GD}{\mbox{$\tilde{G}$}}
\newcommand{\gf}{\mbox{$\gamma_{5}$}}

\newcommand{\Ima}{\mbox{Im}}

\begin{document}

\normalsize
\begin{flushright}{BINP 95-22\\ UQAM-PHE/95-03\\ March 1995}
\end{flushright}
\vspace{0.5cm}
\begin{center}{\Large \bf Electric dipole moment of neutron in the
Kobayashi-Maskawa model with four generations of quarks.}\\
\vspace{1.0cm}
{\bf C. Hamzaoui}\footnote{E-mail: hamzaoui@mercure.phys.uqam.ca}\\
D\'epartement de Physique, Universit\'e du Qu\'ebec \`a Montr\'eal,\\
Case Postale 8888, Succ. Centre-Ville, Montr\'eal, Qu\'ebec, Canada, H3C 3P8.

\vspace{0.7cm}

and\\
\vspace{0.7cm}

{\bf  M.E. Pospelov}\footnote{E-mail: pospelov@inp.nsk.su}\\
Budker Institute of Nuclear Physics, 630090 Novosibirsk, Russia
\vspace{3.5cm}
\end{center}

\begin{abstract}
We show that the existence of a possible fourth heavy generation
of quarks gives rise to a significant enhancement to the neutron
electric dipole moment in comparison with the Standard Model
prediction. The smaller degree of suppression in this case is
linked to the presence of the operators of dimension $\leq$ 6
which enter into the effective Lagrangian with coefficients
proportional to the square of the top quark mass.  Numerically,
the enhancement is mainly associated with chromoelectric dipole
moment of the s quark which appears at three loop level, of the
order $\al_s\al_w^2m_sm_t^2/m^4_w$ from the CP-odd combination
of mixing angles between second, third and fourth generations.
Its value is calculated explicitly in the limit of large masses
of the fourth generation of quarks. The corresponding
contribution to the electric dipole moment of the neutron is
$5\cdot10^{-30} e\cdot cm$ in the most optimistic scenarios
about the values of the Kobayashi-Maskawa matrix elements. The
additive renormalization of $\theta$-term in this model is
estimated as $10^{-13}$.

\end{abstract}

\newpage

\section{Introduction}
The Kobayashi-Maskawa (KM) model looks now as the most natural description
of CP-violation. It describes properly CP-odd phenomena in the decays of
neutral $K$-mesons and predicts extremely tiny CP-odd effects in the
flavour-conserving processes. The current experimental
limit on the electric dipole moment of neutron (EDM) \cite{LG},
\be
d_N/e<10^{-25}\, cm,
\ee
exceeds the realistic Standard Model prediction for this quantity
by seven orders of magnitude. However, this gap between theory and
experiment is of great use for limiting a new CP-violating physics
beyond the Standard Model. The purpose of this work is to consider the
electric dipole moment of neutron in the model with an additional
heavy generation of quarks preserving the same KM origin of
CP-violation.

The reason for introducing an extra heavy generation of quarks
into physics comes from a relatively large mixing in
$B-\bar{B}$ meson system \cite{HSS}. Its existence
is, of course, questionable. However, the set of constraints
on mixing angles and unknown heavy masses can be derived from
the low energy phenomenological data. The analysis \cite{HSS}
shows that the existence of additional heavy quarks with masses
not lighter than $m_t$ is not excluded. We would use the
conclusions developed in the work \cite{HSS} for possible values
of KM matrix elements in terms of the Wolfenstein parameter
$\lambda=|V_{us}|=0.22$.

Let us denote the fourth generation
flavours as (h\,\,\,g). Then the best scenario about a large mixing
between the third and fourth generations consistent with
current experimental data is given by:

\be
|V_{hb}|\sim|V_{tg}|\sim\cal{O}(\lambda)
\label{eq:mix}
\ee
This scenario is quite natural if we assume the masses of h and
g quarks lying below the perturbative unitarity limit of $500$
GeV. Using the unitarity conditions for the KM matrix and
already known values of matrix elements we deduce the following:
\be
|V_{cg}|\sim|V_{hs}|\sim{\cal O}(\lambda^2);\;|V_{hd}|\sim|V_{ug}|\sim{\cal O}
(\lambda^3).
\label{eq:ord}
\ee

The enlarged KM matrix of this model possesses 9 independent
parameters which include six mixing angles and three CP-violating
phases. To avoid the uncertainties of reparametrization, we would
describe all CP-odd flavour-diagonal amplitudes in terms of
imaginary part of three independent quartic combinations of KM
matrix elements:
\begin{eqnarray}
\Ima(V^*_{td}V_{tb}V^*_{cb}V_{cd})\sim{\cal O}(\lambda^7);\;
\Ima(V^*_{td}V_{tb}V^*_{hb}V_{hd})\sim{\cal O}(\lambda^7);\nonumber\\
\Ima(V^*_{ts}V_{tb}V^*_{hb}V_{hs})\sim{\cal O}(\lambda^5)
\label{eq:quart}
\end{eqnarray}
All other rephasing invariants could be reexpressed using
these three combinations and moduli of KM matrix elements only.

\section{Standard Model prediction}

Let us start from the Standard Model (SM) prediction for the
electric dipole moment (EDM) of neutron. The violation of
the CP-symmetry in the Standard Model originates from the sole complex
phase in
the KM matrix. To lowest, quadratic order in the weak interaction
all CP-odd flavour-conserving amplitudes turn to zero trivially.
The point is that in this approximation those amplitudes depend
only on the moduli squared of elements of the KM matrix, so the
result cannot contain the CP-violating phase.  In the next,
quartic order in semi-weak coupling constant $g_w$, the expression
for EDMs of quarks vanishes at two-loop approximation without hard
gluon radiative correction taken into account \cite{Sh}.  It can
be shown, however, that the inclusion of one hard gluon loop
prevents EDMs from identical cancellation.

The values of EDMs to one gluon loop accuracy was calculated first
by Khriplovich \cite{Kh}.  We quote here his result:
\be
d_{d}=e\tilde{\delta}\fr{\als G_F^2m_c^2m_d}{108\pi^5}
\log\fr{m_t^2}{m_b^2}\log^2\fr{m_b^2}{m_c^2}\simeq e\cdot 2 \cdot 10^{-34}cm,
\label{eq:3g}
\ee
where $G_F=\sqrt{2}g_w^2/(8M^2)$ is the Fermi constant and $m_i$ is
the mass of i-flavoured quark. $\tilde{\delta}$ denotes here the
only possible CP-odd invariant of 3 by 3 KM matrix. The analogous
result for $d_u$ is much smaller being proportional to $m_s^2$.

Formula (\ref{eq:3g}) was obtained in the limit of the
effective four-fermion contact interaction which is valid when
all quark masses and characteristic loop momenta are much
smaller than masses of $SU(2)$ gauge bosons. This is not exactly
true in general because the top-quark does not satisfy this demand.
 However, we
could use the expression (\ref{eq:3g}) as a good estimate for
EDM of d-quark in the Standard Model. The reason being is that
$m_t$ enters to (\ref{eq:3g}) under logarithm only and may be
replaced by $m_w$.

The EDM of neutron in Standard Model, however, is much larger than
EDMs of its constituents due to the so called "long
distance" effects \cite{KZ,Gav,McKH}. The most reliable
estimates according the rules of chiral perturbation theory
predicts the EDM of neutron at the level $d_N\simeq
2\cdot 10^{-32}e\cdot cm$. The two orders of magnitude enhancement
here is basically due to a smaller number of closed loops and
bigger factor connected with strong interactions.

Our basic idea is to extend this consideration to the case of KM
model with an additional heavy generation of fermions. The reason
is that instead of one CP-violating phase, the
enlarged variant of this model possesses three CP-odd parameters
and additional flavour's combinations in the amplitudes. It may
cause the contributions to EDMs being proportional to the square
of the mass of the top-quark. As a result, short distance contributions to
EDMs of quarks would be suppressed not by $G_F^2m_c^2$ but rather
by $\al_wG_F$.

In general, the problem of EDM calculation should be divided into
two independent parts. First, we construct a low energy effective
Lagrangian in terms of u, d, s quarks, gluons and external
electromagnetic field. Then, we recalculate this Lagrangian to the
EDM of neutron using all available methods for doing low energy
hadronic physics. The second part is independent from the concrete
model of CP-violation at high energies. It is clear that in our
case the possible enhancement should be associated with effective
operators of low dimension, not bigger than 6. It is easy to
identify all these possible structures. Integrating out c, b, t, g,
h quarks and SU(2) gauge fields, we construct a low energy
effective Lagrangian in the form:
\be
{\cal L}_{eff}(x)=\theta\fr{\als}{8\pi}G^a_{\mu\nu}\GD^a_{\mu\nu}+
i\fr{c_W}{6}g_s^3f^{abc}\GD^a_{\alpha\beta}G^b_{\beta\mu}G^c_{\mu\alpha}+
\sum_i\fr{\tilde{d}_i}{2}g_s\bar{q}_it^a(G^a\s)\gf q_i+
\sum_i\fr{d_i}{2}e\bar{q}_i(F\s)\gf q_i,
\label{eq:Leff}
\ee
where $(F\s)$ denotes $F_{\mu\nu}\gm\gn$ and the summation
is held over light flavours: $i=u,\,d,\,s$.  The first term in
(\ref{eq:Leff}) represents the induced $\theta$-term,
perturbative contribution to the total $\theta$-term of the
theory. The next operator of dimension 6 was introduced
originally by Weinberg \cite{Wein}; the SM value of
corresponding coefficient $c_W$ was calculated in \cite{Posp}.
Other terms in (\ref{eq:Leff}) with dim=5 are the operators of
EDMs and chromoelectric dipole moments (CEDM). The dimension
5 of these operators in some sense is fictitious because of a
chirality flip making coefficients $\tilde{d}_i$ to be
proportional to light masses $m_i$. Therefore, both $d_i$ and
$\tilde{d}_i$ are suppressed by at least two powers of heavy
mass corresponding to the weak interaction scale. Other CP-odd
operators of dimensions higher than 6 are unimportant in our
consideration because they are suppressed by additional powers
of heavy masses.

Some comments should be made at this point. We do not add to
(\ref{eq:Leff}) CP-odd mass operators of quarks $i\bar{q}_i \gf
q_i$ because they could be incorporated to the $\theta$-term by
mean of the chiral rotation.  Other flavour conserving CP-odd
operators of dim=6 built from four quark fields are
suppressed in comparison with CEDMs. The last fact refers to our
concrete model of CP-violation. Because of the V - A character of
the theory, effective operators are originally formulated in terms of
left handed fields.  In its turn, CP-violation may only arise
through the chirality flip which gives extra powers of light mass
and affects the suppression of four quark operators. This last remark
is connected with the so called "axial polyp operators" introduced
originally in
\cite{RGP}:
\be
c_P\bar{q}_i \gm\{G_{\mu\nu},\vec{D}_\nu\}\gf q_i.
\ee
This operator is identical to CEDM one if we assume the equations of
 motion to
be preserved:
\be
i\gm D_\mu q_i=m_i q_i
\label{eq:em}
\ee

Further analysis is recalled to single out leading operators in
(\ref{eq:Leff}) and find corresponding coefficients.

\section{Flavour's structure of ${\cal L}_{eff}$}

We start investigating the coefficients in (\ref{eq:Leff}) by
determining the flavour's
arrangement along a fermion line.
Let us denote by f the Green function of f-flavoured fermion. Then
a CP-odd amplitude for fermionic operators in quartic order in
semi-weak constant could be written in the
following form:
\be
\sum_{j,k,l}i\Ima(V^*_{jf}V_{jk}V^*_{lk}V_{lf})\;fjklf.
\label{eq:fs1}
\ee
For pure gluonic operators the corresponding structure looks as:
\be
\sum_{f,j,k,l}i\Ima(V^*_{jf}V_{jk}V^*_{lk}V_{lf})\;fjkl,
\label{eq:fs2}
\ee
where the cyclic permutation of the kind $fjkl=lfjk=klfj=jklf$ is
allowed.

In three generation formulation of KM model one has the only
source of CP violation which fixes unique flavour structures for
both type of operators \cite{Kh}. In the four family case, the
number of CP violating phases is three. Their concrete choice
could be done in a different ways complicating our analysis.
However, it does not affect the main property responsible for
the cancellation of EDMs at two loops. It is easy to see that
independently on the number of families the expression
(\ref{eq:fs1}) is antisymmetric under the interchange of flavours j and l:
\be
\sum_{j,k,l}i\Ima(V^*_{jf}V_{jk}V^*_{lk}V_{lf})\;fjklf=
\fr{1}{2}\sum_{j,k,l}i\Ima(V^*_{jf}V_{jk}V^*_{lk}V_{lf})\;f(jkl-lkj)f.
\ee
This antisymmetry is sufficient to set to zero both EDM and CEDM
of quark to two-loop approximation \cite{Sh}.

It is useful to classify a variety of CP-odd amplitudes by mean
of dynamical arguments. The enhancement of the short distance
contributions to the ${\cal L}_{eff}$ which we expect to get is
closely related to the fact that all characteristic loop momenta
are comparable with the weak interaction scale. Therefore,
inside the loops, we are legitimate to put all quark masses to zero
except $m_t$, $m_g$ and $m_h$. In other words, inside the loops,
we are able to identify propagators of light quarks:
\be
c=u\equiv U;\,\,\,\, d=s=b\equiv D.
\ee
It should be mentioned here that c and b quarks play a twofold
role. They are considered as heavy quarks at normal hadronic scale
and almost massless inside loops, at the scale of weak
interactions. This requirement immediately leads to the vanishing
of pure gluonic operators as well as EDM and CEDM of u-quark.
Indeed, applying the unitarity condition for the KM matrix
\be
V^*_{du}V_{dk}+V^*_{su}V_{sk}+V^*_{bu}V_{bk}=\delta_{uk}-V^*_{gu}V_{gk}
\ee
we perform the summation over $j$ and $l$ in (\ref{eq:fs1})
explicitly:
\be
\sum_{j,k,l}i\Ima(V^*_{ju}V_{jk}V^*_{lk}V_{lu})\;ujklu=
\sum_{k}i\Ima(V^*_{gu}V_{gk}V^*_{gk}V_{gu})u(Dkg-gkD)u=0.
\ee

For s and d quark operators the situation is quite different. An
analogous procedure for them leads to the following combination:
\be
\fr{1}{2}\sum_k i\Ima(V^*_{tf}V_{tk}V^*_{hk}V_{hf})
f[tkh-hkt+Ukt-tkU+hkU-Ukh]f,
\label{eq:fsD}
\ee
where $f=d,\;s$.

Now we will concentrate on s-quark operators because its mixing
with third and fourth generations is {\em a priori} bigger than
that of d-quark.

Taking the last sum over $k$ we obtain:
\be
\fr{i}{2}\Ima(V^*_{ts}V_{tg}V^*_{hg}V_{hs})
s[t(g-D)h-h(g-D)t+U(g-D)t-t(g-D)U+h(g-D)U-U(g-D)h]s.
\label{eq:fss}
\ee
The rephasing invariant combination of KM matrix elements in
(\ref{eq:fss}) to a good accuracy coincides with that responsible
for CP-odd $B^0_S$ meson mixing:
\be
\Ima(V^*_{ts}V_{tg}V^*_{hg}V_{hs})=-\Ima(V^*_{ts}V_{tb}V^*_{hb}V_{hs}),
\ee
and the resulting expression takes the form:
\be
\fr{i}{2}\Ima(V^*_{ts}V_{tb}V^*_{hb}V_{hs})
s[t(b-g)h-h(b-g)t+U(b-g)t-t(b-g)U+h(b-g)U-U(b-g)h]s.
\label{eq:FSfin}
\ee
 It may be of the order $\lambda^5$ and it makes the electric and
chromoelectric dipole moment of s-quark the most important between
other CP-odd operators.

For the d-quark all considerations presented above are valid with the
only replacement in rephasing invariants:
\be
\Ima(V^*_{ts}V_{tb}V^*_{hb}V_{hs})\longrightarrow
\Ima(V^*_{td}V_{tb}V^*_{hb}V_{hd})\sim {\cal O}(\lambda^7).
\label{eq:l7}
\ee

\section{EDM of neutron}

When the flavour structure is fixed, it is possible to find the
relative meaning of different operators for the EDM of neutron using
order of magnitude estimations for corresponding coefficients.

First we take an electric dipole moment of d-quark. Combining
together all phase space factors, coupling constants and taking
into account (\ref{eq:l7}) we get the following estimate:
\be
d_d\sim \lambda^7 \frac{\alpha_s}{4\pi}
\frac{\alpha_w}{4\pi}\frac{1}{16\pi^2}G_Fm_d \frac{m_t^2}{m_w^2}.
\label{eq:dd}
\ee
In the next
section, we will find that this estimate is reasonable.
Now using a simplest constituent model, we obtain the
corresponding contribution to the EDM of neutron at the level:
\be
d_N\simeq d_d\sim e\cdot 3\cdot10^{-32}cm.
\ee
It turns out that it does not exceed the Standard Model
prediction. The enhancement in comparison with SM value of $d_d$
is just two orders of magnitude instead of $m_t^2/m_c^2\simeq
10^4$. It can be explained rather trivially. In contrast with the SM
result (\ref{eq:3g}), the estimate (\ref{eq:dd}) does not possess
any logarithmic enhancement.  Another origin of this deficiency is
related to a smaller value of $\al_s$ in (\ref{eq:dd}) and smaller
numerical factor in comparison with that of (\ref{eq:3g}).

In the same manner we estimate the chromoelectric dipole moment of
s-quark:
\be
\tilde{d}_s\sim\Ima(V^*_{ts}V_{tb}V^*_{hb}V_{hs}) \frac{\alpha_s}{4\pi}
\frac{\alpha_w}{4\pi}\frac{1}{16\pi^2}G_Fm_s
\frac{m_t^2}{m_w^2}\sim\lambda^5\frac{\alpha_s}{4\pi}
\frac{\alpha_w}{4\pi}\frac{1}{16\pi^2}G_Fm_s \frac{m_t^2}{m_w^2}.
\label{eq:est}
\ee
The renormalization to hadronic scale does not seriously change
this estimate. The recalculation of the contribution of this
quantity to the EDM of
neutron is a separate problem. Here we use the result of
Khatsimovsky, Khriplovich and Zhitnitsky \cite{KKZ} which does not
exhibit any additional suppression of the contribution to NEDM
from this operator:
\be
d_N\simeq -\fr{1}{2}\tilde{d}_s\sim e\cdot 5\cdot 10^{-30} cm.
\ee
This estimate shows that the four-generation formulation of the KM
model in the most optimistic scenario about possible values of
CP-odd rephasing invariants leads to the electric dipole moment of
neutron two orders of magnitude bigger than its SM value. The
electric dipole moment of s-quark appears to be of the same order
of magnitude as $\tilde{d}_s$. However, its contribution to NEDM
seems to be suppressed in comparison with that of CEDM operator
\cite{KKZ}.

It is useful also to estimate values of other terms in
${\cal L}_{eff}$. Both the induced $\theta$-term and Weinberg
operator turn out to be suppressed by the ratio $m_b^2/m_w^2$. It
means also that the characteristic loop momenta could range
widely, from $m_b$ to electroweak scale. Therefore, it is quite
possible that these contributions would match also a large
logarithmic factor so the total suppression would be of order
$m_b^2/m_w^2\log(m_w^2/m_b^2)\sim 1/40$. Thus, our estimate for
the induced value of $\theta$-term is:
\be
\theta\sim\lambda^5\frac{1}{8}\frac{\alpha_s}{4\pi}
\frac{\alpha_w}{4\pi}
\frac{m_t^2m_b^2}{m_w^4}\log(m_w^2/m_b^2)\simeq 3\cdot 10^{-13}.
\label{eq:theta}
\ee
This value is just three orders of magnitude smaller than the
current limit on the total $\theta$-term. Its contribution to the
EDM of neutron could make sense only after specifying a mechanism
of CP-strong puzzle solution.  It is clear that the popular
elimination of $\theta$-dependence due to axions makes the EDM of
neutron unfeasible for (\ref{eq:theta}). There are, however, some
alternative solutions for CP-strong problem which assume the
$\theta$-relaxation at tree level only. This solution makes the
most important the $\theta$-term contribution to the EDM of
neutron.

The last operator of interest is the Weinberg operator. Its
distinguishing feature is the existence without hard gluon
radiative corrections taken into account \cite{Posp}. However its
numerical contribution to EDM in the four-family case is unlikely
to exceed $10^{-33} e.cm$. The additional smallness here is related
to the strong suppression from the renormalization to
hadronic scale.

\section{The limit of infinitely heavy $m_h$ and $m_g$}

We now check if the estimate (\ref{eq:est}) is correct
enough and prove the absence of exact cancellation or additional suppression
at three loop level.

The expression of interest is a three loop integral where quarks
and SU(2) gauge bosons with comparable masses are involved. To
simplify the problem, we impose on masses an artificial condition
\be
m_{h}^2,\,m_{g}^2 \gg m_t^2 \gg m_w^2.
\label{eq:scale}
\ee
which would allow to use an effective scale separation. This
limit in general breaks the perturbative unitarity and forces us
to take into account further electroweak loops. However, it
may be used to estimate the first coefficient of perturbative
series. In our case it brings some important simplification to
the problem of the loop calculation. The most
important contributions are then associated with longitudinal parts of
W-boson's Green functions. Indeed, if the limit (\ref{eq:scale})
is held, all characteristic loop momenta could range between $m_t$
and heaviest masses. From that scale, according to our assumption,
$m_w$ could be regarded as a small mass and $1/m_w^2$ in the
longitudinal parts of W-propagator makes its relative
contribution to the effect to be enhanced in  comparison with that of
Feynman parts. Thus, it is clear that the quantity of interest,
$\tilde{d}_s$, will be proportional to the factor $m_t^2/m_w^4$
and we omit possible contributions of order $1/m_w^2$ and
$1/m_t^2$. In other words, it means that in t'Hooft - Feynman
gauge we take into account only diagrams with charged scalar Higgses
and neglect those with W-bosons. The accuracy of the assumption
(\ref{eq:scale}) for real massive parameters is presumably
about 1/4 if we keep masses of heaviest quarks around
500 $GeV$ and it is sufficient for our purposes.

Two different possibilities of W-bosons attachment to the fermion line are
depicted in Figs. 1a and 1b. These skeletons should be dressed by one hard
gluon loop and external soft gluon leg. Our calculation of the CEDM is based
upon the external field technique proposed by J. Schwinger for QED and then
extended on to the QCD case by Novikov, Shifman, Vainshtein and
Zakharov (see for ex. the review \cite{NSVZ}). It
can surely be applied to this problem because all loop momenta are much
larger than the characteristic hadronic scale.  The technique deals with
the operator
\be
\langle x|P_\mu|y\rangle=\langle
x|iD_\mu|y\rangle=(i\fr{\partial}{\partial
x_{\mu}}+g_s\fr{\lambda^c}{2}A_{\mu}^c(x))\delta^4(x-y)
\label{eq:P}
\ee
where $A_{\mu}^c(x)$ is the gluonic field. Then the quark
propagator taken in the background gluonic field reads as:
\be
\langle vac|Tq^a(x)\bar{q}^b(y)|vac\rangle=\langle
x,a|i(\hat{P}-m)^{-1}|y,b\rangle=\langle
x,a|(\hat{P}-m)\fr{i}{P^2+ig_s/2(G\s)-m^2}|y,b\rangle,
\label{eq:gf}
\ee
where $\Ph\equiv\gm P_\mu$. The field
strength originates here as a result of commutation of two $P$'s:
\be
[\Pm,\Pn]=ig_sG^a_{\mu\nu}\fr{\lambda^a}{2}\equiv ig_sG_{\mu\nu}.
\label{eq:[,]}
\ee

Using this technique, it is easy to demonstrate the vanishing of
CEDM at two loops. This cancellation occurs before the last
integration over a momentum of outer W-boson loop. The part of
the amplitude of interest, Fig.  1a, comprises mass operator
between two propagators antisymmetrized in masses:
\be
\fr{\Ph}
{\Ph^2-m_j^2}\Gh\fr{\Ph}{\Ph^2-m_l^2}-(m_j\,\leftrightarrow\,m_l),
\label{eq:2loop}
\ee
The sum over i and j is performed according the prescription
(\ref{eq:FSfin}).  Here $\Gh$ denotes the mass operator taken in the
background gluonic field. It possesses V - A gamma matrix
structure and allows for the expansion in series of external
field operators of increasing dimension with some invariant
functions depending on $P^2$ as coefficients. The explicit
antisymmetrization in masses in
(\ref{eq:2loop}) leads to the following expression for this
amplitude:
\be
(m_j^2-m_l^2)\fr{\Ph}{(\Ph^2-m_j^2)(\Ph^2-m_l^2)}
[\Gh,(\Ph)^2]\fr{\Ph}{(\Ph^2-m_j^2)(\Ph^2-m_l^2)}
\label{eq:comm}
\ee
The commutator in this expression could be calculated for all
operators entering to $\Gh$. It could be shown that the result of
commutation starts from operators with  several powers of field
strengths or field derivatives.  Therefore, the two-loop amplitudes
cannot induce CEDM (see Ref.\cite{Posp} for details). Clearly, at
three loop accuracy this commutator must be a source of hard gluon
field which compensates some extra dimensions. Another point of hard
gluon attachment is not fixed and all other fermion lines should be
expanded up to the first order in its field. The similar procedure
could be performed over diagrams of the second type.

It is clear that the result of integration may contain some
power of $\log(m_h^2/m_t^2)$ or $\log(m_g^2/m_t^2)$. Naively we
can regard this logarithm as a big parameter and calculate all
diagrams in the "Leading Logarithm Approximation". To this
approximation we believe that $\log(m_h^2/m_t^2)\ll
\fr{1}{2!}\log^2(m_h^2/m_t^2)\ll\fr{1}{3!}\log^3(m_h^2/m_t^2)$.
It is not true for real values of our masses. However to obtain an order
of magnitude, we put all $\fr{1}{n!}\log^n$ equal to 1 in the
final answer. The reason for assuming this artificial condition
is to simplify the set of multi-loop calculation reducing it to
the consequence of factorized integrations.

It turns out that the logarithmic accuracy allows to find all
relevant operators in $\Gh$ expansion and single out leading
diagrams for the rest of amplitude. We start from the smallest
distances determining the internal mass operator $\Gh$ for
diagrams in Fig. 1a and one of mass operators in Fig.  1b. These
distances are associated with the propagation of g and h quarks
respectively. The V - A gamma-matrix structure together with
gauge invariance give three possible operators:
\be
\Gh=\fr{1}{m_w^2}\left(c_0\Ph^3+c_1\gn g_sD_\mu G_{\mu\nu}+
c_2g_s\gm\{P_\nu,G_{\alpha\beta}\}\epsilon_{\mu\nu\alpha\beta}\right)
\fr{1-\gf}{2}\equiv ({\cal O}_0+{\cal O}_1+{\cal O}_2),
\label{eq:gamma}
\ee
The choice of these operators is not unique. We could use another
basis; $\{\Ph,P^2\}$, for example instead of $\Ph^3$, etc. However, the
expansion (\ref{eq:gamma}) appears to be the most convenient. The first
term in this series, ${\cal O}_0$ operator, identically vanishes
being substituted into (\ref{eq:comm}). For the second type of
diagram this operator gives $m_s^3$ being applied to the s-quark
wave function and therefore can be omitted as well. Two other
operators in this series enter here with numerical coefficients
$c_1$ and $c_2$ whith no momentum or mass dependencies. They could
be calculated from the expansion of mass operator in $\Ph$ at
$P^2\ll m^2$.
\be
-i\Gh=-\fr{g_w^2}{2m_w^2}\int\!\fr{d^4q}{(2\pi)^4q^2}\qh\fr{1-\gf}{2}
\fr{1}{\qh-m}\Ph \fr{1}{\qh-m}\Ph\fr{1}{\qh-m}\Ph\fr{1}{\qh-m}\qh
\fr{1-\gf}{2}
\label{eq:expan}
\ee
We used here the W-boson's Green function in the unitary
gauge:
\be
-i\fr{g_{\mu\nu}-q_\mu q_\nu /m_w^2}{q^2-m_w^2}\simeq i\fr{q_\mu q_\nu}
{m_w^2q^2}
\label{eq:prop}
\ee
It is easy to see that the logarithmic divergence of (\ref{eq:expan}),
converting to $\log(m^2/p^2)$ after applying GIM mechanism, is
associated with the ${\cal O}_0$ operator only. The straightforward
calculation leads to the following result:
\be
\fr{c_1}{m_w^2}=-\fr{\al_w}{4\pi}\fr{1}{12};\;\;\;\;
\fr{c_2}{m_w^2}=-\fr{\al_w}{4\pi}\fr{5}{48};\;\;\;\;
\label{eq:c2}
\ee

We are left with two-loop expressions which could give a square of
logarithm in the final answer. This means that the integration
over a momentum flowing at the t quark line should be performed
in last turn as we integrate from small distances to large ones. The
corresponding diagrams and top quark propagation are shown in
Figs. 2a and 2b. The blob represents operator $\Gh$; dashed line
here is the hard gluon propagator.  It is clear that to
logarithmic accuracy, the result of hard gluon loop integration
may be presented as
an effective vertex of W-boson with a fermion line changing the
flavour from t to s.

The position of external gluon field is not indicated in Figs. 2
and all propagators should be taken in the background gluonic
field. At first look, the perturbative expansion breaks the
main advantage of the calculation in the external field - we
have to fix explicitly the gauge of SU(3) field. This problem
could be resolved by dividing the four-potential $A_\mu^a$ into
two parts:
\be
A_\mu^a=(A_\mu^a)_{ext}+a_\mu^a,
\label{eq:Aa}
\ee
where $(A_\mu^a)_{ext}$ is the vacuum field while $a_\mu^a$
denotes the hard gluon field. It is a matter of convenience to
choose an additional term fixing the gauge of $a_\mu^a$ in the
following form (so called background gauge):
\be
-\fr{1}{2}(D_\mu^{ext}a_\mu^a)^2,
\label{eq:fix}
\ee
where $D_\mu^{ext}a_\nu^a=\partial_\mu
a_\nu^a+g_sf^{abc}(A_\mu^b)_{ext}a_\nu^a$.  It is easy to see that
the gauge invariance for $(A_\mu^a)_{ext}$ field is still
preserved. The hard gluon propagator in the external field in
this gauge takes the form \cite{NSVZ}:
\be
\int i\exp\{iqx\}d^4x\langle Ta_\mu^a(x)a_\nu^b(y)\rangle=(\delta^{ab}
\fr{g_{\mu\nu}}{q^2}-2ig_sf^{acb}A^c_\lambda(y)\fr{q_\lambda}{q^4}
+\fr{2g_s}{q^4}f^{acb}G^c_{\mu\nu}+...)\exp\{iqx\},
\label{eq:gp}
\ee
where we have omitted the subscript "ext". This form of propagator
is very useful for our problem and it allows us to calculate CEDM
covariantly without fixing the gauge of external field.

To reduce the number of diagrams the limit of large $N_c$ is
used. In that limit combinations $t^at^at^b;\;\, if^{abc}t^at^c
\simeq \frac{N}{2}t^b$ are much larger than
$t^at^bt^a=\frac{1}{2N}t^b$. It means in particular that the
external chromoelectric field can not be attached to the fermion
line {\em inside} hard gluon loop as it shown at Fig.3. The
accuracy of this approximation is presumably $\fr{1}{N_c^2}=1/9$
and it is within the errors connected with previous
assumptions.

Let us first evaluate the contribution to this vertex from the
operator ${\cal O}_1\sim\gn D_\mu G_{\mu\nu}$ which is a usual
"penguin". By virtue of the equation of motion
\be
D_\mu G^a_{\mu\nu}=-g_s \bar{q}\gn t^aq
\ee
and Fiertz transformation this operator generates a close fermion
loop as it is shown in Fig. 4. The use of this equation is valid
if the characteristic momenta inside this loop could be regarded
smaller than $m_h(m_g)$ which is satisfied to logarithmic
accuracy. Both topologies of W-boson attachment generates a
nonvanishing contribution to CEDM of s-quark. It turns out,
however, that to logarithmic accuracy there is a cancellation
between these two types of diagrams and therefore penguins
operators cannot contribute to the effect at the level of
$\log^2(m_h^2/m_t^2)$.  It is possible to demonstrate that there
is a nonvanishing contribution from these operators beyond this
logarithmic approximation with a dependence of $m_h^2/m_g^2$
parameter.  It requires tideous true multi-loop calculations
which is beyond the scope of our purpose.

A similar cancellation between two topologies occurs when
we substitute ${\cal O}_2$ operator and look for the effective
vertex of W-boson with fermion. This operator, however,
generates additional diagrams which should be taken into account
as well. The first one represents a two loop mass operator
inside W-boson loop (Fig. 5a); the second is the flavour
changing mass operator inside gluonic loop (Fig. 5b). The
crosses here indicate the chirality flips on the fermion line.

The calculation of inner mass operator depicted in Fig. 5a in use of
operator ${\cal O}_2$ is simple. To logarithmic accuracy, it is given
by
\be
M=\fr{c_2}{m_w^2}\fr{\al_sN_c}{4\pi}m_t\log(\fr{m_g^2}{p^2})
g_s(G\s)\fr{1-\gf}{2},
\label{eq:massop}
\ee
where $p$ is the momentum flowing at the fermion line. It ranges
between $m_t$ and the lightest mass of fourth generation quarks.
The summation over flavours annihilates all other structures
without the chirality flip. The only structure which does not vanish
is the operator (\ref{eq:massop}). Let us assume for the moment
that $m_h^2\ll m_g^2$. Then this sum takes the form:
\begin{eqnarray}
[(G\s),\ph]\left(\fr{m_t^2}{p^2(p^2-m_t^2)}\log(\fr{m_g^2}{p^2})-
\fr{m_h^2}{p^2(p^2-m_h^2)}\log(\fr{m_g^2}{m_h^2})+\right.\nonumber\\
\left.\fr{m_h^2}{(p^2-m_t^2)(p^2-m_h^2)}\log(\fr{m_g^2}{m_h^2})\right)
\simeq [(G\s),\ph]\log(\fr{m_h^2}{p^2}),
\end{eqnarray}
where we have neglected a further noncommutativity of momenta
resulting from operators of dimension higher than that of CEDM. It is
very natural that the logarithm is cut off at the lightest
mass between $m_h$ and $m_g$. The last integration is also trivial and
gives the CEDM operator of s-quark with a coefficient
\be
-\Ima(V^*_{ts}V_{tb}V^*_{hb}V_{hs})\fr{G_F}{\sqrt{2}}\frac{\alpha_sN_c}{4\pi}
\frac{1}{16\pi^2}
\frac{c_2m_sm_t^2}{m_w^2}\fr{1}{2!}\log^2(\fr{m_h^2(m_g^2)}{m_t^2}).
\label{eq:CEDM1}
\ee

The last graph given by Fig. 5b contains flavour changing mass
operator. Its value at the incoming momenta $k$, which is much larger than
$m_t$, before the renormalization is given simply by:
\be
M(\kh)=\kh f(k^2)(1-\gf)=\fr{\al_w}{4\pi}\fr{3}{4}\fr{m_t^2}{m_w^2}\kh
\log(\fr{\Lambda^2}{k^2})\fr{1-\gf}{2},
\ee
where $\Lambda$ is the cut-off. The on-mass-shell
renormalization prescription with respect to different masses to
the left and to the right from the mass operator could be found in
Refs. \cite{Sh,Posp}. In our case this prescription looks
trivial because both these masses are negligibly small in
comparison with electroweak scale:
\be
M_r(\kh)=\kh(f(k^2)-f(k^2=0))(1-\gf) =
-\fr{\al_w}{4\pi}\fr{3}{4}\fr{m_t^2}{m_w^2}\kh
\log(\fr{k^2}{m_t^2})\fr{1-\gf}{2}.
\ee
The remaining integration over $k$ ranging between $m_t$ and
$m_h(m_g)$ is equivalent to that performed
several steps earlier, in (\ref{eq:massop}). The corresponding
contribution to CEDM of s-quark differs from (\ref{eq:CEDM1}) by
the factor -3. Combining these two numbers together and taking
into account the value of the coefficient $c_2$, we obtain the
final answer for the CEDM of s-quark for  very heavy fourth
generation, to double logarithmic accuracy and in the limit of
large $N_c$:
\be
\tilde{d}_s=-\Ima(V^*_{ts}V_{tb}V^*_{hb}V_{hs}) \fr{G}{\sqrt{2}}m_s
\frac{\alpha_s\alpha_w}{(4\pi)^4}\frac{5N_c}{12}
\frac{m_t^2}{m_w^2}\fr{1}{2!}\log^2(\fr{m_h^2(m_g^2)}{m_t^2}).
\label{eq:answer}
\ee
Substitution of $N_c=3$ and
$\fr{1}{2!}\log^2(\fr{m_h^2(m_g^2)}{m_t^2})\sim 1$ to this formula yields the
estimation of CEDM close to (\ref{eq:est}).

\section{Conclusions}

We have demonstrated a new interesting feature of the model with
four generations of quarks incorporated into the same KM
mechanism.  The enhancement of neutron EDM in comparison with SM
prediction comes from small distance effects which provide a
regular factor of order $m_t^2/m_c^2$. The resulting value of
EDM in this model, however, is just two orders of magnitude
larger than corresponding SM value.  The reason for that is in
the numerical importance of large distance contribution in SM
which is two orders of magnitude bigger than that coming from EDMs
or CEDMs of quarks.

The estimation of relevant operators in the effective
Lagrangian performed in this work allows one to consider the
influence of the fourth generation on other low-energy
CP-violating observables such as T-odd form-factors of heavy
nuclei. This problem deserves special consideration.

\begin{flushleft}{\Large\bf Acknowledgements}\end{flushleft}

We thank C. Burgess, V. Chernyak, G. Couture, V. Khatsimovsky,
I. Khriplovich, A. Vainshtein and A. Yelkhovsky for helpful
stimulating discussions.

\newpage
\vspace{3in}

\begin{center}{\large\bf Figures}\end{center}
\vspace{1.3in}
\begin{center}{Fig. 1a\hspace{3in}Fig. 1b}\end{center}
\vspace{1.3in}
\begin{center}{Fig. 2a\hspace{3in}Fig. 2b}\end{center}
\vspace{1.3in}
\begin{center}{Fig. 3}\end{center}
\vspace{1.3in}
\begin{center}{Fig. 4}\end{center}
\vspace{1.3in}
\begin{center}{Fig. 5a\hspace{3in}Fig. 5b}\end{center}

\end{document}